\documentclass[twocolumn]{aastex63}

\def\be{\begin{equation}}
\def\ee{\end{equation}}
\def\ba{\begin{eqnarray}}
\def\ea{\end{eqnarray}}
\usepackage{mathptmx}
\usepackage{indentfirst}
\usepackage{ulem}
\usepackage{graphicx}	
\usepackage{amsmath}	
\usepackage{amssymb}	
\usepackage{amsmath}
\usepackage{listings}
\usepackage{color}
\usepackage{url}

\graphicspath{{./}}

\received{XXX}
\revised{XXX}
\accepted{XXX}

\submitjournal{ApJ}

\shorttitle{Multitracer technique for skew spectrum}
\shortauthors{Dai \& Xia}

\begin{document}

\title{Constraints on primordial non-Gaussianity using multitracer technique for skew spectrum}
\correspondingauthor{Jun-Qing Xia}
\author{Ji-Ping Dai}
\email{daijp@mail.bnu.edu.cn}
\affiliation{Department of Astronomy, Beijing Normal University, Beijing 100875, China}
\author{Jun-Qing Xia}
\email{xiajq@bnu.edu.cn}
\affiliation{Department of Astronomy, Beijing Normal University, Beijing 100875, China}

\begin{abstract}
Extracting the bispectrum information from the large scale structure observations is challenging due to the complex models and the computational costs to measure the signal and its covariance. Recently, the skew spectrum was proposed to access parts of the bispectrum information with a more effective way and has been confirmed it can provide complementary information to that enclosed in the power spectrum
measurements. In this work, we  generalize  the theory to apply the multitracer technique and explore its ability to constrain the local type primordial non-Gaussianity. Using the spectra and their covariance estimated from $N$-body simulations, we find the multitracer approach is effective to reduce the cosmic variance noise. The $1\sigma$  marginalized errors for $b_1^2A_s, n_s$ and $f_{\rm NL}^{\rm loc}$ are reduced by 50\%, 52\% and 73\% comparing with the results using only power spectrum obtained from a single tracer. It indicate that both the skew spectrum and the multitracer technique are useful to constrain the primordial non-Gaussianity with the forthcoming wide-field galaxy surveys.
\end{abstract}


\section{Introduction}
\label{sec:intro}

The standard inflationary paradigm predicts a flat universe perturbed by nearly Gaussian scale-invariant primordial perturbations. These predictions have been extensively probed by the increasingly precise measurements of the cosmic microwave background (CMB) \citep{aghanim2018planck}. Different from the CMB, the large scale structure (LSS) contains 3-dimensional distribution information of the galaxies on large scales, which are caused by the nonlinear evolution due to the gravitational instability. The upcoming wide-field galaxy surveys, such as DESI \citep{aghamousa2016desi}, EUCLID \citep{amendola2018cosmology} and LSST \citep{Abell:2009aa}, can provide complementary information on the origin of our Universe and its late-time evolution.

The traditional method to extract the cosmological information from the LSS is measuring the 2-point correlation function or the power spectrum in Fourier space. However, due to the late-time gravitational instability, the galaxy distribution at low redshift is highly non-Gaussian, even for Gaussian initial conditions. To obtain more information form the same surveys, higher-order statistic will be an intuitionistic method to apply, such as 3-point correlation function and bispectrum \citep{Matarrese:1997sk,Verde:1998zr, scoccimarro2000bispectrum, sefusatti2006cosmology, hoffmann2015measuring}. Actually, the bispectrum  has been measured using galaxy survey data \citep{scoccimarro2001bispectrum, verde20022df, marin2013wigglez, gil2015power} and has been proven useful to break degeneracies among cosmological parameters which arise from considering the power spectrum alone \citep{Gil-Marin:2014baa,Gil-Marin:2016wya}. With the forthcoming surveys, the higher-order statistics can reach a much larger signal-to-noise ratio, and provide a wealth of information.

However, due to the complicated triangle configurations and orientations, it requires significant computational efforts to measure the bispectrum signal and its covariance, and is more challenging to compare the theoretical models with measurements. To bypass these problems, there are several proxy statistics  proposed to compress the bispectrum to a pseudo-power spectrum, which only depend on one wavenumber but contain some of the information enclosed in the bispectrum. One of the approaches is the integrated bispectrum proposed by \citet{chiang2014position}, which is generated by cross correlating the position-dependent power
spectrum with the mean overdensity of the corresponding subvolume. This measurement contains the bispectrum information on squeezed configuration and has been detected using real data \citep{chiang2015position}. The order method is the skew spectrum which was first studied in CMB \citep{cooray2001squared, munshi2010new}, and then adopted to LSS \citep{pratten2012non, schmittfull2015near, munshi2017integrated, dizgah2019capturing, Dai:2020adm}. The skew spectrum is obtained by  cross correlating the square of a field with the field itself, and has been proven that it is an effective method to access complementary information to that enclosed in the power spectrum measurements using $N$-body simulations.

Primordial non-Gaussianity (PNG) is one of the most important fingerprints of inflation and can be used to discriminate between the vast array of inflationary scenarios. Currently, the most strict constraints have been achieved by the CMB temperature anisotropies and polarizations, and the amplitudes of the  local, equilateral, and orthogonal types are: $f_{\mathrm{NL}}^{\text {loc }}=-0.9 \pm 5.1 ; f_{\mathrm{NL}}^{\text {equil }}=-26 \pm 47 ; \text { and } f_{\mathrm{NL}}^{\text {ortho }}=-38 \pm 24$ at $1\sigma$ statistical significance \citep{Akrami:2019izv}. However, such strict constraints on $f_{\rm NL}$ have not been obtained from LSS measurements, although the halo bias can be greatly affected by relatively small values of $f_{\rm NL}$ as shown by \citet{grossi2009large} using numerical simulations. The latest constraints on $f_{\mathrm{NL}}^{\text {loc }}$ was from the BOSS quasar samples, and the result is $-51<f_{\mathrm{NL}}^{\text {loc }}<21$  at 95\% confidence level \citep{Castorina:2019wmr}. \citet{Dai:2020adm} has shown that with the measurements of skew spectrum, the $1\sigma$ marginalized error for $f_{\mathrm{NL}}^{\text {loc}}$ can be reduced by 44\% although with a large smoothing filter, which suggests the skew spectrum is an effect method to constrain the PNG without significant computational costs.

Another import issue is that the clustering analysis at large scales where the PNG signal is most significant is limited by the cosmic variance (CV). A possible method to reduce the CV is the multitracer technique \citep{Seljak:2008xr, Slosar:2008ta, Ferramacho:2014pua, Yamauchi:2014ioa, Fonseca:2015laa, Yamauchi:2016wuc} which can significant improve the statistical errors using different biased tracers. For two different tracers $\delta_i$ and $\delta_j$, we can obtain 4 cross skew spectra from $\delta_i^2$$\times$ $\delta_j$, $\delta_j^2 \times \delta_i$, $\delta_i\delta_j \times \delta_j$, $\delta_i\delta_j \times \delta_i$, and there is only one cross power spectrum from $\delta_i$$\times$ $\delta_j$. We expect that we can obtain a tighter constraints on $f_{\mathrm{NL}}^{\text {loc}}$ using multitracer technique for skew spectrum.

In this paper, we build on our previous work \citep{Dai:2020adm} and include the multitracer technique. We simply divide our simulated halo catalog into two parts, and then calculate the cross power spectra and skew spectra to find the extra information the multitracer technique can give us. The rest of the paper is organized as follows. In Sec. \ref{sec:meth} we briefly review the full expression for the skew spectrum including both the primordial non-Gaussianity and the late-time non-Gaussianity, then we extent our theory to apply the multi-tracer technique. In Sec. \ref{sec:sim} we show how we divide our $N$-body simulation catalog and derive the covariance of the power spectra and skew spectra. In Sec. \ref{sec:res} we list the constraint results and conclude in Sec. \ref{sec:con}. We also derive the Poisson shot noise contributions to the galaxy power spectrum and skew spectrum in App. \ref{app}.

\section{Methodology}
\label{sec:meth}

\subsection{General expression for matter skew spectrum}
To begin with, we define the matter overdensity field $\delta({\vec x})=\delta\rho(\vec x)/\bar{\rho}$ where $\bar \rho$ is the spatial average of the matter density. We can write the 3-point correlation function as
\be
\label{eq:3-cf}
\xi^{(3)}(\vec{x_{1}},\vec{x_{2}},\vec x_{3})=\left\langle\delta\left(\vec{x}_{1}\right) \delta\left(\vec{x}_{2}\right) \delta\left(\vec{x}_{3}\right)\right\rangle.
\ee
This is a well known statistic to extract the extra information not  present in the power spectrum. However, as we explained before, it is challenging to measure form LSS data.

To simplify the 3-point correlation function, we can assume $\vec x_{3}$ in Eq. (\ref{eq:3-cf}) is located at the same point as $\vec x_{1}$, which means we cross correlate the square of the field $\delta^2$ with the $\delta$ field itself. This statistic is called the skew correlation function, and due to the cosmological
principle, it depends only on the magnitude of the separation vector:
\begin{equation}
\xi^{(s)}(x_{12}) \equiv \xi^{(3)}(\vec{x_{1}},\vec{x_{1}},\vec x_{2})= \xi^{(s)}(|\vec x_{1}-\vec x_{2}|) .
\end{equation}
Following \citet{Dai:2020adm}, we can perform the Fourier transformation of this equation to obtain the matter skew spectrum:
\be
\begin{aligned}
\label{eq:gen}
P_m^{(s)}(k)&=\int \frac{d^{3} \vec{q}}{(2 \pi)^{3}}B_m(k,q,|\vec{q}-\vec{k}|) \\
&=\int_{-1}^1 d\mu \int \frac{d{q}}{(2 \pi)^2}q^2B_m(k,q,\alpha(\mu)),
\end{aligned}
\ee
where $B_m(k,q,|\vec{q}-\vec{k}|)$ is the bispectrum of the overdensity field, $\mu=\vec{k}\cdot\vec{q}/kq$ and $\alpha=\sqrt{q^2+k^2-2\mu k q}$ ensures that the wavenumbers correspond to the three sides of a triangle.

In order to calculate the matter skew spectrum, we need to explicit the matter bispectrum $B_m$ whose main contributions are from primordial perturbations $B_{m,I}$ and gravitational instability $B_{m,G}$. Here we discuss these two effects separately. This part has been widely studied in \citet{pratten2012non, schmittfull2015near, chan2017assessment, munshi2017integrated, dizgah2019capturing, Dai:2020adm}. Since this paper focuses on the quasi-linear scales, we only consider the leading order contributions in the following analysis.

First, Let us begin with the local type primordial non-Gaussianity which is the main target of this paper. The Bardeen's curvature perturbation during the matter era is given by \citep{salopek1990nonlinear, Gangui:1993tt, verde2001tests, komatsu2001acoustic},
\be
\Phi(\vec{x})=\Phi_{G}(\vec{x})+f_{\rm NL}^{\rm{loc}} \left[\Phi_{G}^{2}(\vec{x})-\left\langle\Phi_{G}^{2}(\vec{x})\right\rangle\right],
\ee
where $\Phi_{G}(\vec{x})$ is a Gaussian field.

To characterize the matter bispectrum, we need to relate the linear density fluctuations with the curvature perturbations. In Fourier space, it can be written as,
\be
\delta(k)^{(1)}=M(k,a)\Phi(k);~ M(k,a) =\frac{2k^2 T(k) D(a)}{3\Omega_{m}H_{0}^{2}},
\ee
where $a$ is the scale factor, $H_0$ is the Hubble constant, $\Omega_m$ is the current matter energy density parameter, $T(k)$ is the matter transfer function and $D(a)$ is the growth factor. It allows us to write the matter bispectrum from primordial perturbations as
\be
B_{m,I}(k_1,k_2,k_3)=M(k_1)M(k_2)M(k_3)B_{\Phi}(k_1,k_2,k_3)\,
\ee
where $B_{\Phi}(k_1,k_2,k_3)$ is the leading order contribution to the  curvature field bispectrum, and it can be expressed as
\be
B_{\Phi}\simeq 2f_{\rm NL}^{\rm{loc}}\left[ P_{\Phi}(k_1)P_{\Phi}(k_2)+\rm cyc. \right],
\ee
where $P_{\Phi}(k)=\left\langle\Phi(k)\Phi^{*}(k)\right\rangle$ is the primordial spectrum.

Even for Gaussian initial conditions, our Universe is highly non-Gaussian due to the late-time non-linear gravitational evolution. Using perturbation theory, the matter  density fluctuations can be expressed as a series of corrections to the linear solution $\delta({\vec{k}})^{(1)}$ (e.g. \citep{bernardeau2002large})
\be
\delta({\vec{k}})=\delta({\vec{k}})^{(1)}+\delta({\vec{k}})^{(2)}+\delta({\vec{k}})^{(3)}+\ldots,
\ee
here we only keep the first two order, and $\delta({\vec{k}})^{(2)}$ is given by
\be
\delta({\vec{k}})^{(2)}=\int \mathrm{d}^{3} \vec q_{1} \mathrm{d}^{3} \vec q_{2} \delta_{D}\left(\vec{k}-\vec{q}_{12}\right) F_{2}\left(\vec{q}_{1}, \vec{q}_{2}\right) \delta({\vec{q}_{1}})^{(1)} \delta({\vec{q}_{2}})^{(1)},
\ee
where $\delta_D$ is the Dirac delta function and $F_2(\vec{q}_{1}, \vec{q}_{2})$ is the known second-order kernel of standard perturbation theory
\be
F_{2}\left(\boldsymbol{q}_{1}, \boldsymbol{q}_{2}\right)=\frac{5}{7}+\frac{x}{2}\left(\frac{q_{1}}{q_{2}}+\frac{q_{2}}{q_{1}}\right)+\frac{2}{7} x^{2},
\label{eq:2OPTkernel}
\ee
with $x \equiv {\vec{q}}_{1} \cdot {\vec{q}}_{2}/q_1q_2$. Then the bispectrum generated by the gravitational instability at leading order is given by
\be
B_{m,G}\left(k_{1}, k_{2}, k_{3}\right)=2 F_{2}\left(\vec{k}_{1}, \vec{k}_{2}\right) P_{m,L}\left(k_{1}\right) P_{m,L}\left(k_{2}\right)+\mathrm{cyc.},
\ee
where $P_{m,L}\left(k\right)$ is the linear matter power spectrum.
The general  expression for matter skew spectrum is
\be
P_{m}^{(s)}(k)=\int_{-1}^1 d\mu \int \frac{d{q}}{(2 \pi)^2}q^2\left[B_{m,I}(k,q,\alpha)+B_{m,G}(k,q,\alpha)\right].
\ee

\subsection{Galaxy skew spectra and power spectra for multi-tracers}
What we actually observe are galaxies and they are biased tracers of the dark matter field. In this paper, we use a simple prescription
in Eulerian space, where the galaxy overdensity is expanded in terms of the matter overdensity and the traceless part of the tidal tensor. Up to the second order, we have (e.g. \citep{desjacques2018large})
\be
\delta_{g}(\boldsymbol{x}) \simeq b_{1} \delta(\boldsymbol{x})+\frac{1}{2} b_{2} \delta^{2}(\boldsymbol{x})+
\frac{1}{2}b_{K^2}\left[\left(\frac{\partial_{i} \partial_{j}}{\partial^{2}}-\frac{1}{3} \delta_{i j}\right)\delta(\boldsymbol{x})\right]^2~,
\ee
where $b_1$, $b_2$ are the linear and non-linear bias and $b_{K^2}$ is the non-local tidal shear bias. As shown in \citet{Dai:2020adm}, the effect of $b_{K^2}$ to the final results is not significant, so we neglect the non-local term in the following analysis.

In Fourier space, the galaxy overdensity is given by
\be
\delta_{g}(\vec{k}) \simeq b_1 \delta(\vec{k})+\frac{1}{2}b_2\int \mathrm{d}^{3} {\vec q} \delta({\vec q})\delta{({\vec k-\vec q})}.
\ee
For a single tracer, the galaxy bispectrum at leading order can be easily expressed as
\be
\begin{aligned}
B_{g,\rm 1T}\left(k_{1}, k_{2}, k_{3}\right)=&b_{1}^{3} \left[ B_{m,I}\left(k_{1}, k_{2}, k_{3}\right)+B_{m,G}\left(k_{1}, k_{2}, k_{3}\right)\right]\\
&+b_{1}^2b_2\left[P_{m,L}\left(k_{1}\right) P_{m,L}\left(k_{2}\right)+\rm{cyc.}\right],
\end{aligned}
\ee
and then the galaxy skew spectrum is given by
\be
P_{g,\rm 1T}^{(s)}(k)=\int_{-1}^1 d\mu \int \frac{d{q}}{(2 \pi)^2}q^2B_{g,\rm 1T}(k,q,\alpha).
\ee

The situation gets more complicated when we consider two tracers which have different bias parameters: $b_1^{[1]}$, $b_2^{[1]}$ for the first tracer and $b_1^{[2]}$, $b_2^{[2]}$ for the second tracer. For example, we cross correlate the  square of the first tracer $(\delta_g^{[1]})^2$ with the second tracer $\delta_g^{[2]}$, and the skew correlation function is given by
\be
\xi^{(s)}(x_{12})=\left\langle\delta_g^{[1]}\left(\vec{x}_{1}\right) \delta_g^{[1]}\left(\vec{x}_{1}\right) \delta_g^{[2]}\left(\vec{x}_{2}\right)\right\rangle.
\ee
The effect of the linear bias is straightforward, which can be written as $(b_1^{[1]})^2b_1^{[2]}\left\langle\delta\left(\vec{x}_{1}\right) \delta\left(\vec{x}_{1}\right) \delta\left(\vec{x}_{2}\right)\right\rangle$. After Fourier transformation, the galaxy skew spectrum due to the linear bias is
\be
\begin{aligned}
P_{g,\rm 2T}^{(s)}(k)|_{\rm LB}=&\int_{-1}^1 d\mu \int \frac{d{q}}{(2 \pi)^2}q^2 (b_1^{[1]})^2b_1^{[2]}\times  \\
&\left[ B_{m,I}\left(k,q,\alpha\right)+B_{m,G}\left(k,q,\alpha\right)\right].
\end{aligned}
\ee
The contribution of the non-linear bias to the correlation function is
\be
\begin{aligned}
\xi^{(s)}(x_{12})|_{\rm NLB}=&\frac{1}{2}(b_1^{[1]})^2b_2^{[2]}\left\langle\delta\left(\vec{x}_{1}\right) \delta\left(\vec{x}_{1}\right) \delta^2\left(\vec{x}_{2}\right)\right\rangle \\
&+b_1^{[1]}b_1^{[2]}b_2^{[1]}\left\langle\delta^2\left(\vec{x}_{1}\right) \delta\left(\vec{x}_{1}\right) \delta\left(\vec{x}_{2}\right)\right\rangle ,
\end{aligned}
\ee
and the corresponding skew spectrum is
\be
\begin{aligned}
P_{g,\rm 2T}^{(s)}(k)|_{\rm NLB}=&\int_{-1}^1 d\mu \int \frac{d{q}}{(2 \pi)^2}q^2
\left\{(b_1^{[1]})^2b_2^{[2]}P_{m,L}(q)P_{m,L}(\alpha)\right.  \\
&\left.+b_1^{[1]}b_1^{[2]}b_2^{[1]}[P_{m,L}(k)P_{m,L}(q)+P_{m,L}(k)P_{m,L}(\alpha)]\right\}.
\end{aligned}
\ee

To sum up, when considering two different tracers, we can obtain six different skew spectra. We use the subscript $(11,2)$ to express the cross correlation spectrum of the  square of the first tracer $(\delta^{[1]})^2$ with the second tracer $\delta^{[2]}$. The full expression of the six skew spectra are:

\begin{widetext}
\begin{eqnarray}
P_{g,(11,1)}^{(s)}(k)&=&\int_{-1}^1 d\mu \int \frac{d{q}}{(2 \pi)^2}q^2\left\{(b_1^{[1]})^3
\left[ B_{m,I}\left(k,q,\alpha\right)+B_{m,G}\left(k,q,\alpha\right)\right]+(b_1^{[1]})^2b_2^{[1]}[P_{m,L}(k)P_{m,L}(q)+{\rm cyc.}]\right\} , \\
P_{g,(22,2)}^{(s)}(k)&=&\int_{-1}^1 d\mu \int \frac{d{q}}{(2 \pi)^2}q^2\left\{(b_1^{[2]})^3
\left[ B_{m,I}\left(k,q,\alpha\right)+B_{m,G}\left(k,q,\alpha\right)\right]+(b_1^{[2]})^2b_2^{[2]}[P_{m,L}(k)P_{m,L}(q)+{\rm cyc.}]\right\} , \\
P_{g,(11,2)}^{(s)}(k)&=&\int_{-1}^1 d\mu \int \frac{d{q}}{(2 \pi)^2}q^2\left\{(b_1^{[1]})^2b_1^{[2]}
\left[ B_{m,I}\left(k,q,\alpha\right)+B_{m,G}\left(k,q,\alpha\right)\right]\right. \\
&&\left.+(b_1^{[1]})^2b_2^{[2]}P_{m,L}(q)P_{m,L}(\alpha)  +b_1^{[1]}b_1^{[2]}b_2^{[1]}[P_{m,L}(k)P_{m,L}(q)+P_{m,L}(k)P_{m,L}(\alpha)]\right\} , \nonumber \\
P_{g,(22,1)}^{(s)}(k)&=&\int_{-1}^1 d\mu \int \frac{d{q}}{(2 \pi)^2}q^2\left\{(b_1^{[2]})^2b_1^{[1]}
\left[ B_{m,I}\left(k,q,\alpha\right)+B_{m,G}\left(k,q,\alpha\right)\right]\right. \\
&&\left.+(b_1^{[2]})^2b_2^{[1]}P_{m,L}(q)P_{m,L}(\alpha)  +b_1^{[1]}b_1^{[2]}b_2^{[2]}[P_{m,L}(k)P_{m,L}(q)+P_{m,L}(k)P_{m,L}(\alpha)]\right\} , \nonumber \\
P_{g,(12,1)}^{(s)}(k)&=&\int_{-1}^1 d\mu \int \frac{d{q}}{(2 \pi)^2}q^2\left\{(b_1^{[1]})^2b_1^{[2]}
\left[ B_{m,I}\left(k,q,\alpha\right)+B_{m,G}\left(k,q,\alpha\right)\right]\right. \\
&&\left.+(b_1^{[1]})^2b_2^{[2]}P_{m,L}(k)P_{m,L}(q)  +b_1^{[1]}b_1^{[2]}b_2^{[1]}[P_{m,L}(k)P_{m,L}(\alpha)+P_{m,L}(q)P_{m,L}(\alpha)]\right\} , \nonumber \\
P_{g,(12,2)}^{(s)}(k)&=&\int_{-1}^1 d\mu \int \frac{d{q}}{(2 \pi)^2}q^2\left\{(b_1^{[2]})^2b_1^{[1]}
\left[ B_{m,I}\left(k,q,\alpha\right)+B_{m,G}\left(k,q,\alpha\right)\right]\right. \\
&&\left.+(b_1^{[2]})^2b_2^{[1]}P_{m,L}(k)P_{m,L}(q)  +b_1^{[1]}b_1^{[2]}b_2^{[2]}[P_{m,L}(k)P_{m,L}(\alpha)+P_{m,L}(q)P_{m,L}(\alpha)]\right\} .\nonumber
\end{eqnarray}
\end{widetext}

Finally it is necessary to review the galaxy power spectra for multi-tracers. Since we only focus on $k<0.1~h\rm Mpc^{-1}$, it is sufficient that we only consider the leading order of the power spectrum. There are three power spectra for two different tracers, which are
\ba
P_{g,(1,1)}(k)= (\tilde b^{[1]})^2P_{m,L}(k), \\
P_{g,(2,2)}(k)= (\tilde b^{[2]})^2P_{m,L}(k), \\
P_{g,(1,2)}(k)= \tilde b^{[1]}\tilde b^{[2]}P_{m,L}(k),
\ea
where the tilde above the bias parameters means the galaxy power spectrum can be greatly affected by relatively small values of $f_{\rm NL}^{\rm loc}$ via the large-scale bias \citep{dalal2008imprints, grossi2009large, wagner2010n,mcdonald2008primordial, matarrese2008effect, sefusatti2009constraining, Dai:2019tjh}, the relationship between $b_1$ and $\tilde b_1$ is given by
\be
\frac{\tilde b_1-b_1}{b_1-1}=2 f_{\rm NL}^{\rm loc} \frac{\delta_{c}}{M(k,z)} q~,
\ee
where $\delta_c \simeq 1.686$ is the threshold for collapse and the correction $q=0.75$ is calibrated from $N$-body simulations \citep{mcdonald2008primordial}

\subsection{Shot noise}

Due to the discrete distribution of galaxies, both power spectrum and skew spectrum have additional stochasticity contributions. In this work, we consider the Poisson model and the number density of the tracers is given by
\be
n(\vec x) = \sum_i \delta_D(\vec x-\vec x_i),
\ee
the discrete  density contrast is defined as
\be
\delta_g = \frac{n(\vec x)}{\bar n} - 1,
\ee
here $\bar n \equiv \left\langle n(\vec x) \right\rangle$ is the mean number density. \citet{Chan:2016ehg} derived the Poisson shot noise of the 2-point and 3-point functions detailedly. Following their work, we derive the shot noise contributions to power spectrum and skew spectrum in App. \ref{app}. The results are listed below, where we use $S(k), S^{(s)}(k)$ to express the shot noise of power spectrum and skew spectrum, respectively.
\ba
S_{\rm (1,1)} &=& \frac{1}{\bar n_1}, \\
S_{\rm (2,2)} &=& \frac{1}{\bar n_2}, \\
S_{\rm (1,1)} &=& 0, \\
S^{(s)}_{\rm (11,1)}&=&\int \frac{d^{3} \vec{q}}{(2 \pi)^{3}} \left[\frac{1}{\bar n_1}\left( P_g(k)+ P_g(q)+P_g(\alpha) \right)+\frac{1}{\bar n_1^2}\right],\\
S^{(s)}_{\rm (22,2)}&=&\int \frac{d^{3} \vec{q}}{(2 \pi)^{3}} \left[\frac{1}{\bar n_2}\left( P_g(k)+ P_g(q)+P_g(\alpha) \right)+\frac{1}{\bar n_2^2}\right],\\
S^{(s)}_{\rm (11,2)}&=&\int \frac{d^{3} \vec{q}}{(2 \pi)^{3}} \frac{1}{\bar n_1}P_{g,(1,2)}(k),\\
S^{(s)}_{\rm (22,1)}&=&\int \frac{d^{3} \vec{q}}{(2 \pi)^{3}} \frac{1}{\bar n_2}P_{g,(1,2)}(k),\\
S^{(s)}_{\rm (12,1)}&=&\int \frac{d^{3} \vec{q}}{(2 \pi)^{3}} \frac{1}{\bar n_1}P_{g,(1,2)}(q),\\
S^{(s)}_{\rm (12,2)}&=&\int \frac{d^{3} \vec{q}}{(2 \pi)^{3}} \frac{1}{\bar n_2}P_{g,(1,2)}(q).
\ea

\subsection{Smoothing}
Even if we truncate the wavenumber range at quasi-linear scales, the galaxy skew spectrum still contains the highly non-linear information due to the integral over $\vec q$. However, the  second-order kernel $F_2$ is only valid on quasi-linear scales, and is expected to fail in non-linear regime. To overcome this problem, there are several fitting formulae of $F_2$ using $N$-body simulations to derive a more reliable expression for the bispectrum \citep{scoccimarro2001fitting, gil2012improved}. However these formulae are only valid in a specific $k$ range.
For simplicity, in our analysis we apply a large smoothing
filter to the field to suppress the small scale modes. By doing this, we may lose some non-linear information, but we can have better analytical control. If the results show the skew spectrum with a large smoothing
filter can improve the constraints, it indicates that using a more sophisticated modelling of the gravitational instability kernel, the analysis could further lift the remaining degeneracies.

In this paper, we use a top-hat windows function whose Fourier transform is
\be
W_{R}(k)=\frac{3 \sin (k R)}{k^{3} R^{3}}-\frac{3 \cos (k R)}{k^{2} R^{2}},
\ee
where $R$ is the radius of the smoothing filter. Then the smoothed power spectra and skew spectra become
\ba
P_{g,R}(k) &=& P_g(k) W^2_R(k) \\
P_{g,R}^{(s)}(k) &=& \int \frac{d^{3} \boldsymbol{q}}{(2 \pi)^{3}}B_g(k,q,\alpha)W_R(k)W_R(q)W_R(\alpha).
\ea

\section{Simulations}
\label{sec:sim}

The frequently used method to seek for the information that the multitracer technique can give us is the Fisher matrix analysis, which is less computationally intensive. However, due to the high correlation between power spectrum and skew spectrum, and the complex properties of the skew spectrum covariance, we resort to numerically computed covariance from a suite of simulations. This is not as fast and simple as a Fisher matrix analysis and requires access to large simulations, but the results will be more reliable.

In our analysis, we use 1000 realizations from the \textsc{Quijote} simulations \footnote{https://github.com/franciscovillaescusa/Quijote-simulations} \citep{villaescusa2019quijote}. The cosmological parameters are: $\Omega_{ m}=0.3175, \Omega_{b}=0.049, h=0.6711, n_s=0.9624, \sigma_8=0.834, M_\nu=0.0~{\rm eV}$, and $f_{\rm NL}^{\rm loc}=0$, which are in good agreement with the latest Planck results \citep{aghanim2018planck}. The simulations were performed with the TreePM code $\textsc{Gadget-III}$, an improved version of $\textsc{Gadget-II}$ \citep{springel2005cosmological}. All the simulations have $512^3$ particles in a box with cosmological volume of 1$(h^{-1}\rm Gpc)^3$. Details of the simulations can be found in \citep{villaescusa2019quijote}.

To study the multitracer technique, we use the halo catalogues which  were identified using the Friends-of-Friends algorithm \citep{davis1985evolution} with linking length $b=0.2$ at $z=0$. We divide each catalog into two parts whose halo mass ranges are: $[2.5\times10^{13}, 1\times10^{14}]~h^{-1}M_{\odot}$ and $[1.3\times10^{13}, 2.5\times10^{13}]~h^{-1}M_{\odot}$, and there are 163000 and 206000 halos on average. Hereafter, we call them T1 and T2 respectively.

It is worth noticing that these simulations have Gaussian initial condition. Thus the constraint results for $f_{\rm NL}^{\rm loc}$ should be consistent with 0 and the error-bars can reflect the constraint ability using different combinations of the power spectra and the skew spectra.

\begin{figure*}[htb]
	\centering
    \includegraphics[width=0.49\linewidth]{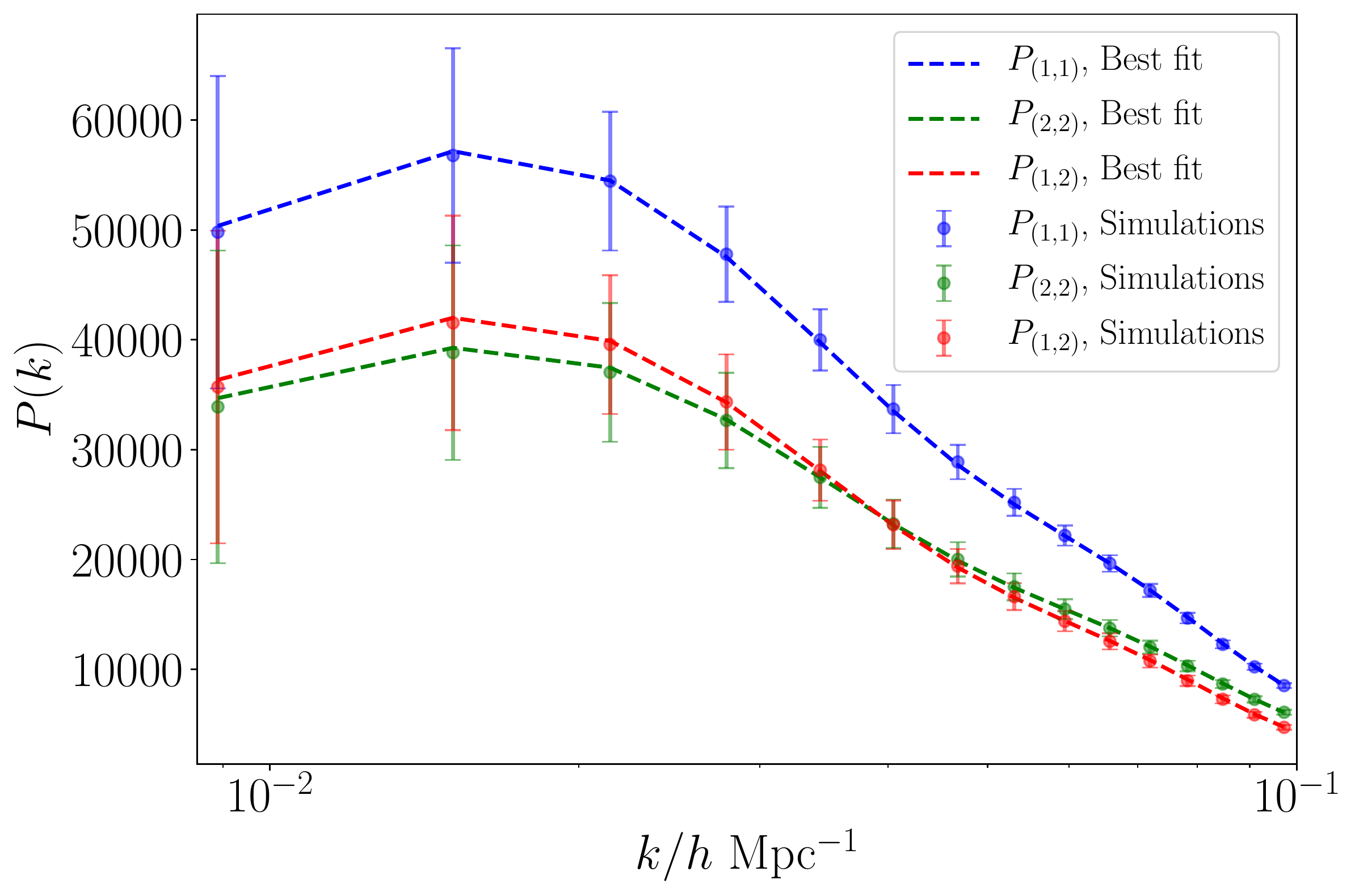}
    \includegraphics[width=0.49\linewidth]{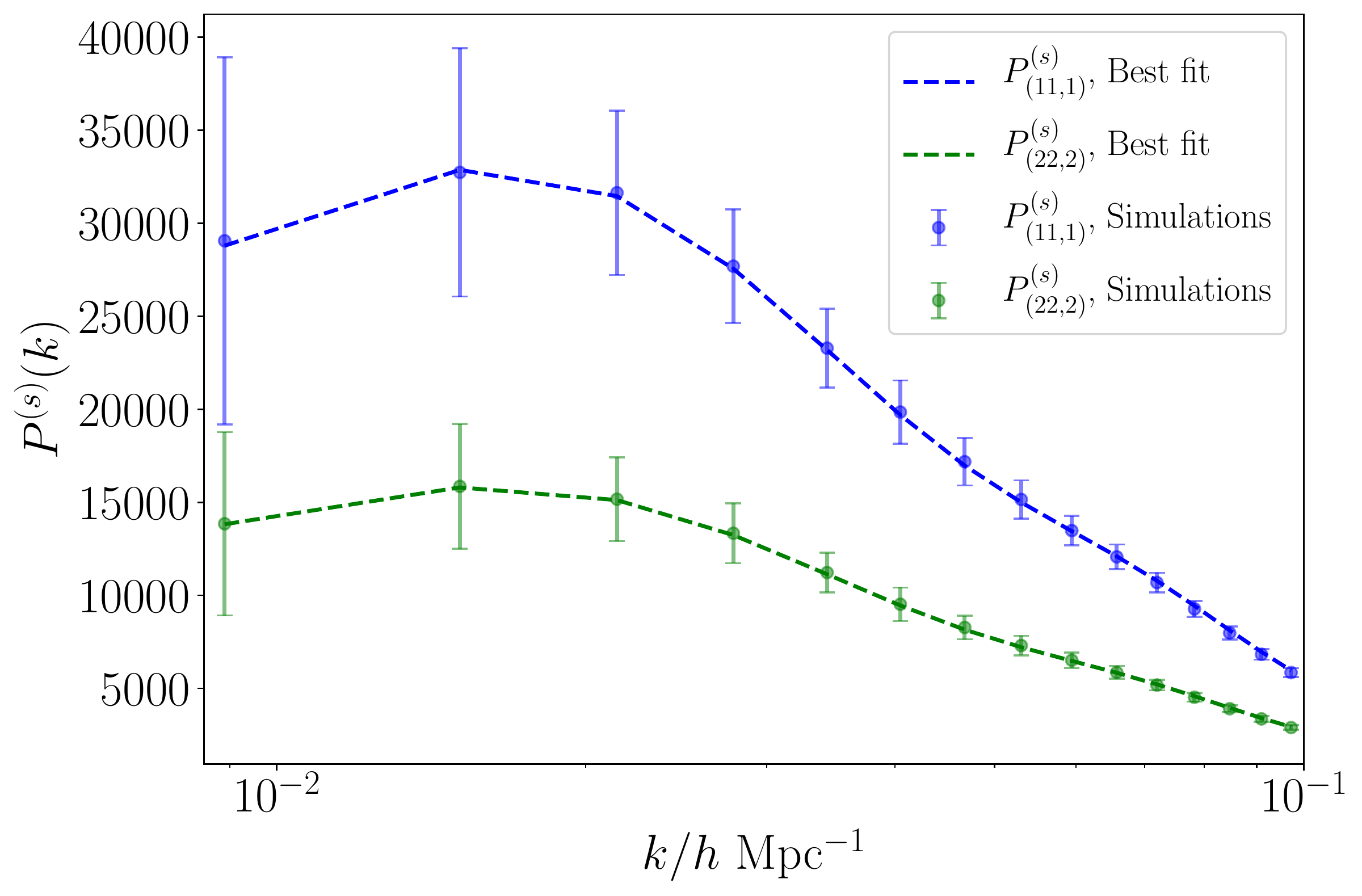}
    \includegraphics[width=0.49\linewidth]{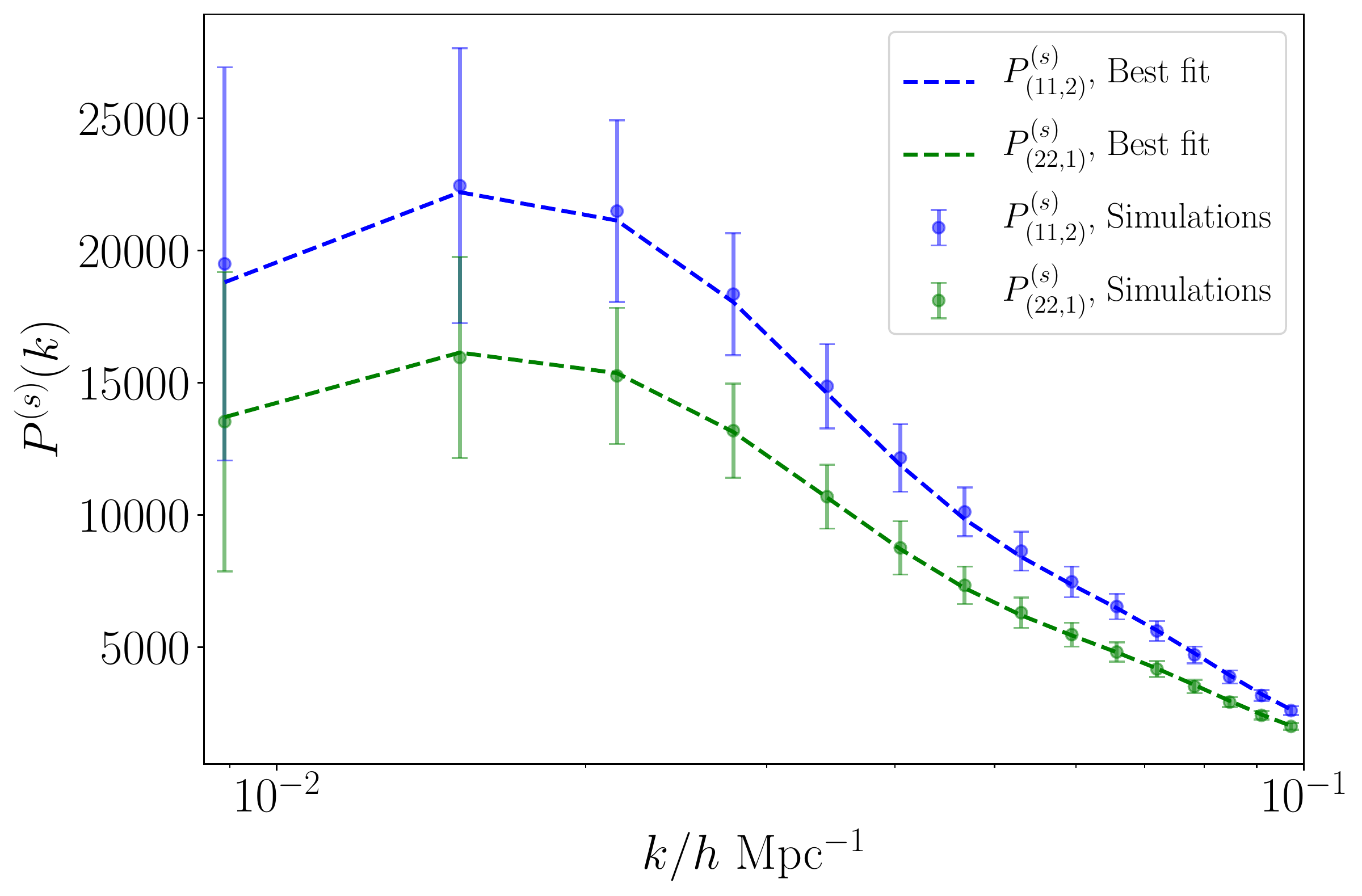}
    \includegraphics[width=0.49\linewidth]{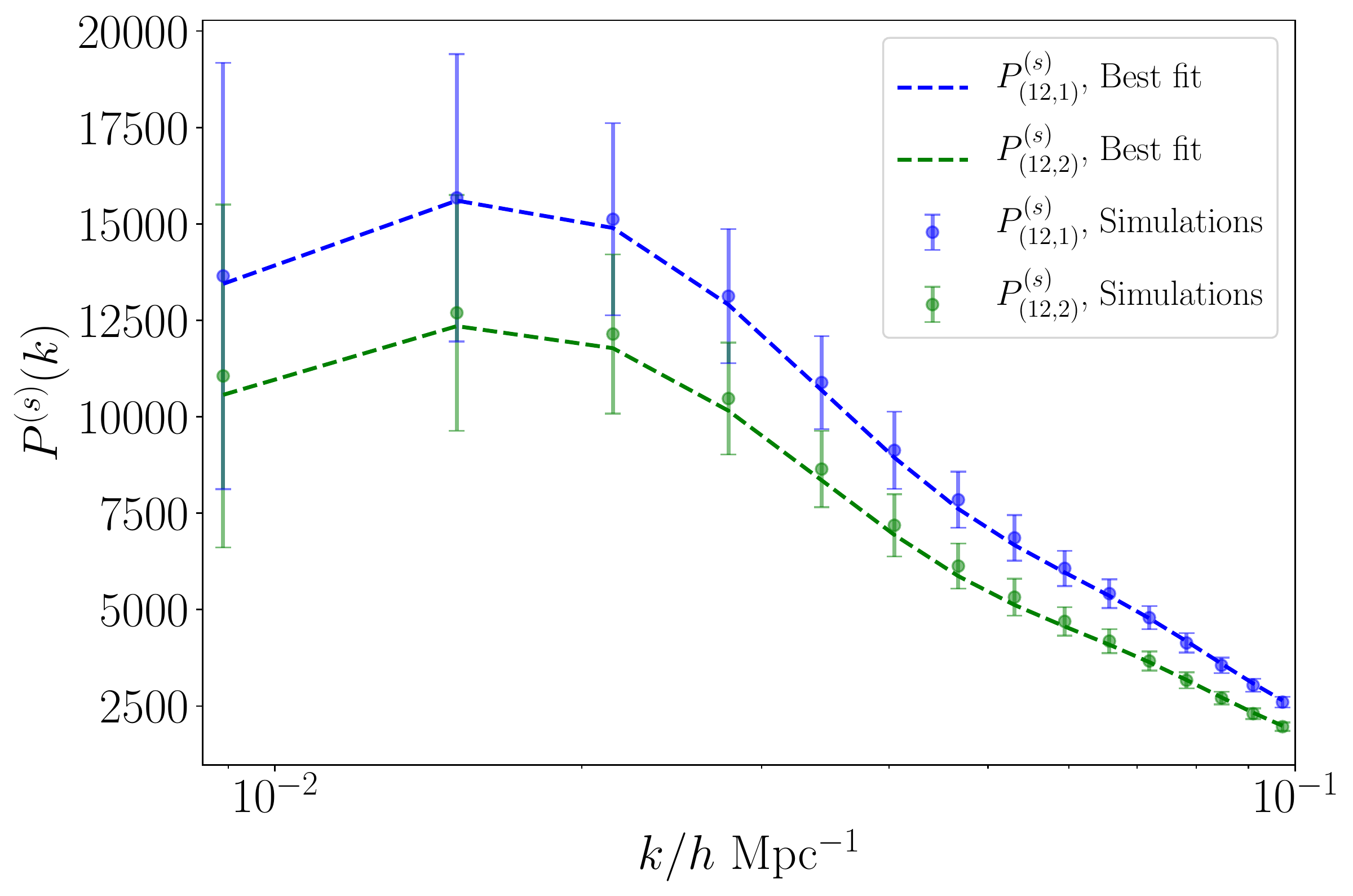}
	\caption{The measured power spectra and skew spectra for multi-tracers (dots and error bars are the average values and the standard deviations of the 1000 realizations), together with the best-fit theoretical models (dashed lines). The upper left panel shows the power spectra $P_{(1,1)}, P_{(2,2)}, P_{(1,2)}$; the upper right panel shows the auto skew spectra $P^{(s)}_{(11,1)}$, $P^{(s)}_{(22,2)}$; the lower left panel shows the cross skew spectra $P^{(s)}_{(11,2)}$ and $P^{(s)}_{(22,1)}$; the lower right panel shows the cross skew spectra $P^{(s)}_{(12,1)}$ and $P^{(s)}_{(12,2)}$. 1 and 2 stand for  T1 and  T2 of our halo catalog.}
	\label{fig:spectra}
\end{figure*}

We use the routine provided in $\textsc{Pylians}$ \footnote{https://github.com/franciscovillaescusa/Pylians} to calculate the cross spectrum of the squared field $\delta^2_h(\vec x)$ and $\delta_h(\vec x)$ field itself. Before squaring the density field, we apply a top-hat smoothing filter with $R = 20h^{-1}\rm Mpc$. In Fig. \ref{fig:spectra} we plot the real space power spectra and skew spectra for multi-tracers obtained from the simulations. The data points are the average results of the 1000 realizations and the error bars are the standard deviations of the spectra at a specific $k$. We also show the theoretical predictions for the best-fit parameters (see details in Sec. \ref{sec:res}). The results show with this smoothing choice, the standard perturbation theory is sufficient to describe the skew spectra.

\begin{figure}[htb]
	\centering
    \includegraphics[width=1.1\linewidth]{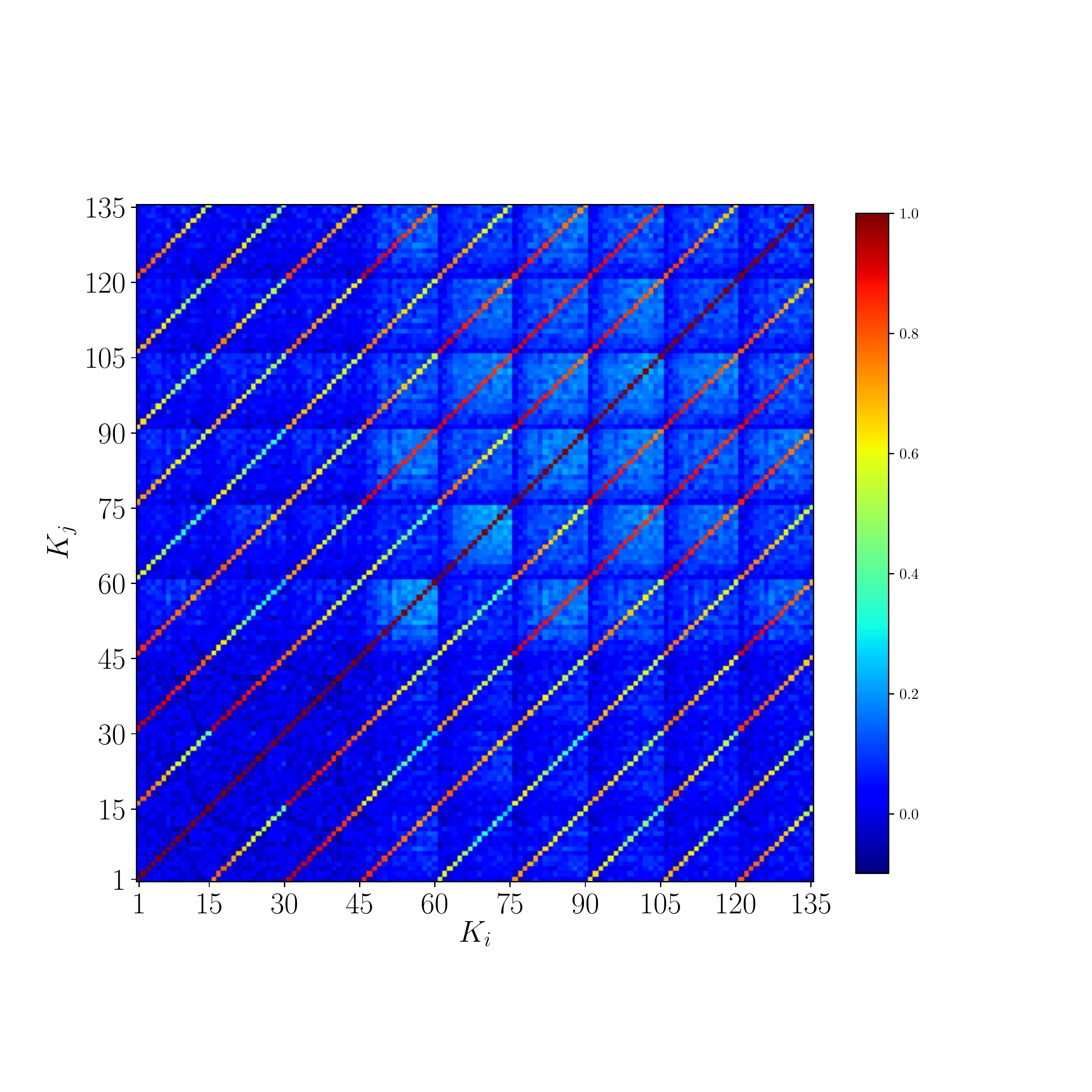}
	\caption{The correlation matrix of $P^{(p+s)}(K_i)$}
	\label{fig:cov}
\end{figure}

Before being able to perform a joint analysis using the power spectra and the skew spectra, we need to evaluate the covariance of these quantities. Since we use a large smoothing filter, only the quasi-linear scales are useful. In our analysis, we use the wavenumber range $k=[0.0089, 0.1]~h\rm Mpc^{-1}$, and there are 15 $k$ bins uniformly spaced in log $k$. We arrange $P_{(1,1)}, P_{(2,2)}, P_{(1,2)}$, $P^{(s)}_{(11,1)}, P^{(s)}_{(22,2)}, P^{(s)}_{(12,1)}, P^{(s)}_{(12,2)}, P^{(s)}_{(22,1)}, P^{(s)}_{(11,2)}$ into a ``data" vector $P^{(p+s)}(K_i)$ (i=1,...,135). In Fig. \ref{fig:cov} we plot the correlation matrix of $P^{(p+s)}(K_i)$, which is defined as
\be
\label{eq:cov}
\frac{C^*_{K_i,K_j}}{\sqrt{C^*_{K_i,K_i}C^*_{K_j,K_j}}},
\ee
where $C^*_{K_i,K_j}$ is the estimated covariance of $P^{(p+s)}_h(K_i)$. We can find that the different $k$ modes are  weakly correlated even for the skew spectra. As \citet{hartlap2007your} pointed, the inverse the covariance matrix is a biased estimator and can be correlated by introducing a Hartlap factor,
\be
 C^{-1} = \frac{n-p-2}{n-1} (C^*)^{-1},
\ee
where $n = 1000$ is the number of independent observations and $p$ is the dimensionality of our data.

\section{Constraint Results}
\label{sec:res}

In our analysis, we consider one of the 1000 realisations as our mock Universe, and with the covariance form simulations we can constrain the cosmological parameters by fitting the power spectra and the skew spectra for multi-tracers. We use four different combinations of the spectra. First we combine T1 T2 and treat it as a single tracer, and constrain the parameters using its power spectrum alone and power spectrum together with skew spectrum. Then we turn to use the multitracer technique and also use the power spectra alone and power spectra together with skew spectra.

We modify the the public software $\textsc{CosmoMC}$ \footnote{http://cosmologist.info/cosmomc/} \citep{lewis2002cosmological}, a Markov Chain Monte Carlo (MCMC) code to perform joint Bayesian parameter inference. A simple $\chi^2$ is used for  parameter fitting in our analysis:
\be
\chi^2=\left[\hat P(K_i)-P(K_i)\right]C^{-1}_{K_i,K_j}\left[\hat P(K_j)-P(K_j)\right]^T,
\label{eq:chisq}
\ee
where $\hat P$ and $P$ represent the model spectra and the measured spectra, and $C^{-1}_{K_i,K_j}$ is their covariance after Hartlap correction. The best-fit parameters can be obtained by finding the minimal of $\chi^2$ and the confidence regions are defined by the surfaces of constant $\Delta \chi^2=\chi^2-\chi_{\min}^2$, where $\chi_{\min}^2$ is the minimal value of $\chi^2$.

We start by determining the bias parameters of T1 and T2 with the other fiducial cosmological parameters fixed. Since the strong degeneracies between biases and the other parameters, this step can be used to check the validity of our theoretical prediction without adding too many variables. Using all the spectra for multi-tracers, the constraints are listed in Tab. \ref{tab:bias}. We find $b_1^{[1]} = 1.451\pm 0.013, b_2^{[1]} = -0.714\pm0.026$ for T1 and $b_1^{[2]} = 1.193\pm0.016, b_2^{[2]} = -0.784\pm0.028$ for T2. The relationship between $b_1$ and $b_2$ is also  consistent with the fitting formula in \citet{lazeyras2016precision, desjacques2018large}
\be
b_2=0.412-2.143 b_{1}+0.929\left(b_{1}\right)^{2}+0.008\left(b_{1}\right)^{3}.
\ee
Using the best-fit bias parameters, we plot the theoretical models with dashed lines in Fig. \ref{fig:spectra}. It shows our theory can accurately predict the measurements at linear scales.

\begin{table}[t]
	\caption{The best-fit values of the bias parameters and  their marginalized  1$\sigma$ errors.}
	\begin{tabular}{c c c c}
    \hline
    \hline
    $b^{[1]}_1$&$b^{[1]}_2$&$b^{[2]}_1$&$b^{[2]}_2$\\
    \hline
    $1.451\pm 0.013$ & $-0.714\pm0.026$ & $1.193\pm0.016$ & $-0.784\pm0.028$\\
    \hline
	\end{tabular}
	\centering
	\label{tab:bias}
\end{table}

\begin{figure}[b]
	\centering
    \includegraphics[width=1.1\linewidth]{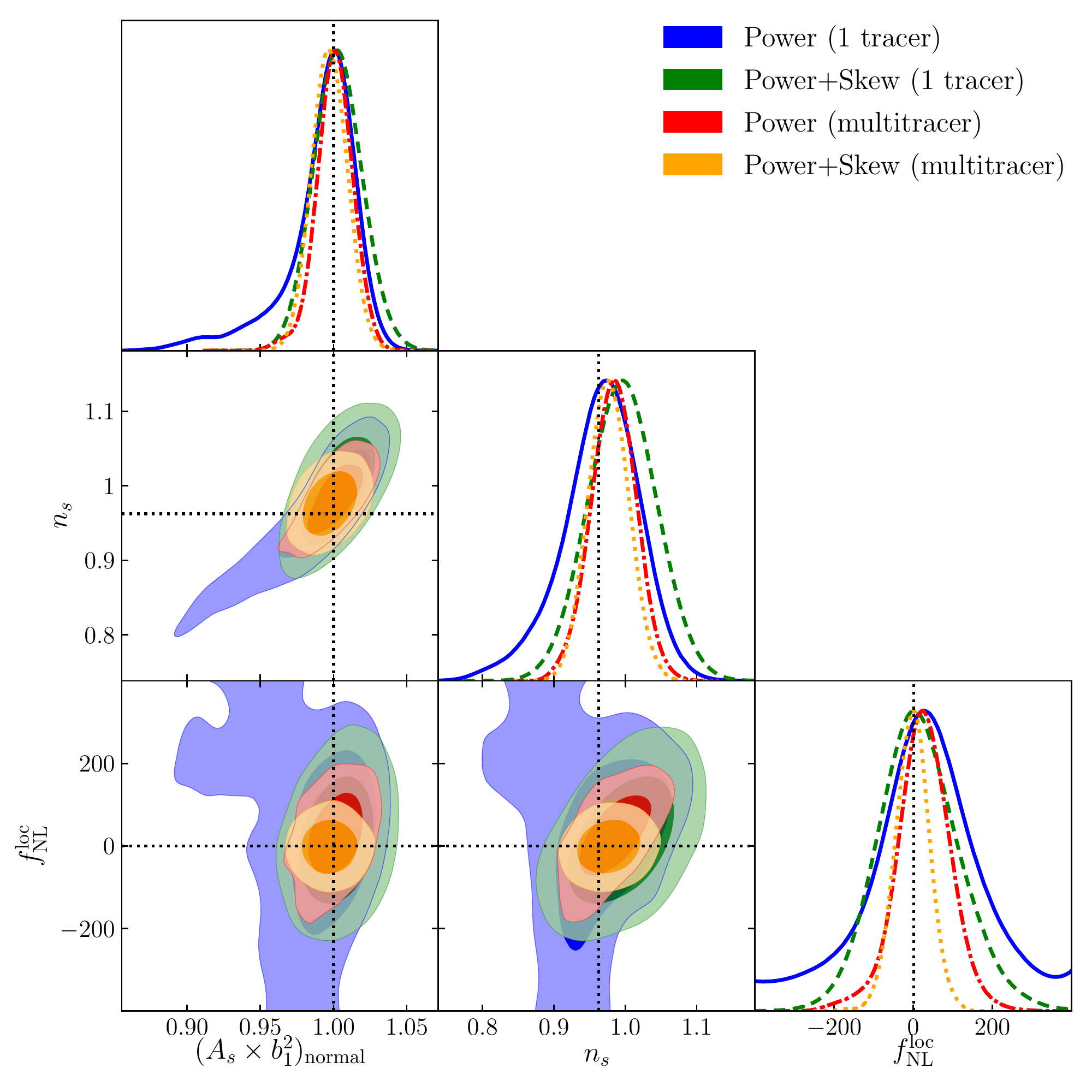}
	\caption{Marginalized two-dimensional distributions and posterior distributions for normalized $b_1^2A_s$, $n_s$, and $f_{\rm NL}^{\rm loc}$. These constraints are obtained from power spectrum (blue) and power spectrum together with skew spectrum (green) using the combination of T1 and T2, and power spectra (red) and power spectra + skew spectra (yellow) with multitracer approach. The black dotted lines are their input  values used in the $\textsc{Quijote}$ simulations}
	\label{fig:d2}
\end{figure}

Now we turn to constrain the cosmological and bias parameters simultaneously to investigate the extra information by using the multitracer technique. The  parameterization we use is
\be
P = \{A_s, n_s, f_{\rm NL}^{\rm loc}, b^{[1]}_1, b^{[1]}_2, b^{[2]}_1, b^{[2]}_2\} ,
\ee
where $A_s$ and $n_s$ are the amplitude and spectral index of the primordial spectrum. The other parameters are fixed at their fiducial values. In Fig. \ref{fig:d2} and Tab. \ref{tab:param} we show our constraint results. Since $A_s, b_1$ and $b_2$ are highly correlated, we construct a new variable $(b_1^2A_s)_{\rm normal} \equiv b_1^2A_s/(b_1^2A_s)_{\rm fid}$. For the multitracer approach, we define this value as the average result of the two tracers. From the results, we find the constriants are consistent with the fiducial values.

\begin{table*}[htb]
	\caption{The best-fit results of normalized $b_1^2A_s$, $n_s$ and $f_{\rm NL}^{\rm loc}$, together with their  marginalized  1$\sigma$ errors.}
	\begin{tabular}{c c c c c}
    \hline
    \hline
    parameters&power (1 tracer)&power + skew (1 tracer) &power (multitracer)&power + skew (multitracer)\\
    \hline
    $(b_1^2A_s)_{\rm normal}$ & $0.997\pm0.024$ & $1.003\pm0.017$ & $1.001\pm0.014$ & $0.997\pm0.012$\\
    $n_s$ & $0.969\pm0.058$ & $0.985\pm0.046$ & $0.982\pm0.034$ & $0.975\pm0.028$\\
    $f_{\rm NL}^{\rm loc}$ & $19.4\pm156.4$ & $10.5\pm99.1$ & $15.2\pm72.4$ & $-4.8\pm42.2$\\
    \hline
	\end{tabular}
	\centering
	\label{tab:param}
\end{table*}

First, we only use the power spectrum and combine T1 and T2, the  marginalized 2-D contours are shown in blue in Fig. \ref{fig:d2}, and the constraint results for $(b_1^2A_s)_{\rm normal}, n_s$ and $f_{\rm NL}^{\rm loc}$ are $0.997\pm0.024, 0.969\pm0.058, 19.4\pm156.4$ (68\% C.L.). The constraints get tighter when we include the skew spectrum, which is already pointed in \citet{Dai:2020adm}. In this analysis, the  addition of the skew spectrum to the power spectrum
yields a reduction of the errors by 29\%, 21\% and 37\% for $(b_1^2A_s)_{\rm normal}, n_s$ and $f_{\rm NL}^{\rm loc}$, respectively. The results are consistent with the conclusion in \citet{Dai:2020adm}.

Then we turn to consider the multitracer technique. When we only use the power spectra, the constraint errors are markedly shrunken, and the results are $1.001\pm0.014, 0.982\pm0.034, 15.2\pm72.4$ (68\% C.L.) for $(b_1^2A_s)_{\rm normal}, n_s$ and $f_{\rm NL}^{\rm loc}$. The constraints are reduced by 42\%, 41\% and 54\% comparing with the single tracer case. It shows the multitracer technique can effectively reduce the cosmic variance, thus the amplitude parameters like $b_1^2A_s$ and $f_{\rm NL}^{\rm loc}$ are better constrained. Due to the degeneracies between cosmological parameters, the errors of the other parameters will also be shrunken. Finally, we use all the power spectra and skew spectra for multitracer approach and the results are shown in yellow in Fig. \ref{fig:d2}, which are $0.997\pm0.012, 0.975\pm0.028, -4.8\pm42.2$ (68\% C.L.) for $(b_1^2A_s)_{\rm normal}, n_s$ and $f_{\rm NL}^{\rm loc}$. Comparing with the results using power spectra for multi-tracers, the $1\sigma$ marginalized errors are reduced by 14\%, 18\% and 42\%. This reduction is due to the extra information that the skew spectra contained. When considering the information from both skew spectra and multitracer technique, i.e., comparing with the results obtained from the power spectrum for a single tracer, the errors are shrunk by  50\%, 52\% and 73\%. Both the skew spectrum and the multitracer technique are effective methods to constrain on  primordial non-Gaussianity.

\section{Conclusions}
\label{sec:con}

In this paper, we mainly discuss the potential power of the multitracer technique for the skew spectrum as a possible probe of the local type primordial non-Gaussianity. The skew spectrum is estimated by cross correlating the squared field $\delta^2(\vec x)$ with the $\delta(\vec x)$ field itself.  Computationally, measuring the skew spectrum is equivalent to a power spectrum estimation, but the skew spectrum contains parts of the 3-point  clustering information, which can be used to further reduce the parameter degeneracies present at the level of the power spectrum. To apply the multitracer technique, we first review the formula of the galaxy skew spectrum which has contributions from primordial non-Gaussianity, gravitational instability and galaxy (halo) bias, and then generalize the theory to multi-tracers to predict both the signals and the shot noise contributions.

Since the high correlation between power spectrum and skew spectrum and the complex properties of the covariance, we do not apply the frequently used Fisher  matrix analysis. Instead, we estimate the  covariance from a suite of simulations and constrain the parameters using a joint Bayesian parameter inference. Our method is not as fast as a  Fisher matrix analysis, but the results are more reliable.

We divide the simulated halo catalog into two parts, which have comparable samples, and then estimate the spectra and their covariance to perform the joint constraints. For comparison, we also constrain the parameters using the whole halos. The results show that by applying the skew spectra and multitracer technique, the $1\sigma$  marginalized errors for  $ (b_1^2A_s)_{\rm normal}, n_s$ and $f_{\rm NL}^{\rm loc}$ are reduced by 50\%, 52\% and 73\%. With the forthcoming generation of wide-field galaxy surveys, the use of the skew spectrum and multitracer technique can offer us a powerful and fast way to constrain the primordial non-Gussianity.

However, due to the large smoothing filter ($20h^{-1}\rm Mpc$) adopted in our analysis, the skew spectrum only contains the linear scale information. If we can obtain a more sophisticated modelling of the gravitational instability kernel using simulations, the analysis can be pushed to smaller scales and  further lifting the remaining degeneracies. We leave this exploration to future work.

\section*{Acknowledgements}
This work is supported by the National Science Foundation of China under grants No. U1931202, 11633001, and 11690023, and the National Key R\&D Program of China No. 2017YFA0402600. We acknowledge the use of the Quijote simulations https://github.com/franciscovillaescusa/Quijote-simulations.


\appendix
\section{Poisson shot noise of power spectrum and skew spectrum}
\label{app}

Firstly, for a single tracer, the 2-point correlation function of the discrete galaxy field is
\be
\xi^{(2)}_{g}\left(\vec{x}_{1}, \vec{x}_{2}\right)=\left\langle\delta_{g}\left(\vec{x}_{1}\right) \delta_{g}\left(\vec{x}_{2}\right)\right\rangle=\frac{1}{{\bar n}^{2}}\left\langle n\left(\vec{x}_{1}\right) n\left(\vec{x}_{2}\right)\right\rangle-1 ,
\ee
where
\be
\label{eq:a2}
\left\langle n\left(\vec{x}_{1}\right) n\left(\vec{x}_{2}\right)\right\rangle = \left\langle \sum_i\delta_D (\vec x_1 - \vec x_i) \delta_D (\vec x_2 - \vec x_i) \right\rangle + \left\langle \sum_{i,j}\delta_D (\vec x_1 - \vec x_i) \delta_D (\vec x_2 - \vec x_j) \right\rangle  = \delta_D(\vec x_1 - \vec x_2) \bar n +{\bar n}^2[1+\xi^{(2)}(\vec{x}_{1}, \vec{x}_{2})].
\ee
Here we need to consider the case when two points are the same, and if the points are different, it can be modeled by the smooth correlation function $\xi^{(2)}$. Thus we can express $\xi^{(2)}_{g}$ as
\be
\xi^{(2)}_{g}\left(\vec{x}_{1}, \vec{x}_{2}\right) = \xi^{(2)}\left(\vec{x}_{1}, \vec{x}_{2}\right)+\frac{1}{\bar n} \delta_D(\vec{x}_{1}- \vec{x}_{2}).
\ee
After Fourier transforming, the galaxy power spectrum is given by,
\be
P_{g,\rm measured}(k) = P_g(k) + \frac{1}{\bar n}.
\ee
The shot noise contribution to the power spectrum of a single tracer is $S_{\rm 1T}(k) = 1/\bar n$. When we consider two different tracers, the first term on the right side of Eq. (\ref{eq:a2}) vanish, so we have $S_{\rm 2T}(k) = 0$.

To calculate the shot noise of skew spectrum, we begin with the 3-point correlation function,
\be
\xi^{(3)}(\vec x_1, \vec x_2, \vec x_3) = \left\langle\delta_{g}\left(\vec{x}_{1}\right) \delta_{g}\left(\vec{x}_{2}\right)\delta_{g}\left(\vec{x}_{3}\right)\right\rangle = \frac{1}{{\bar n}^3} \left\langle n\left(\vec{x}_{1}\right) n\left(\vec{x}_{2}\right)n\left(\vec{x}_{3}\right)\right\rangle - \left[\frac{1}{{\bar n}^{2}}\left\langle n\left(\vec{x}_{1}\right) n\left(\vec{x}_{2}\right)\right\rangle + 2{\rm cyc.} \right] +2 ,
\ee
where the three-point correlator of $n$ is
\be
\label{eq:a6}
\begin{aligned}
\left\langle n\left(\vec{x}_{1}\right) n\left(\vec{x}_{2}\right) n\left(\vec{x}_{3}\right)\right\rangle=&\left\langle\sum_{i} \delta_{{D}}\left(\vec{x}_{1}-\vec{x}_{i}\right) \delta_{{D}}\left(\vec{x}_{2}-\vec{x}_{i}\right) \delta_{{D}}\left(\vec{x}_{3}-\vec{x}_{i}\right)\right\rangle+\left[\left\langle\sum_{i, j} \delta_{{D}}\left(\vec{x}_{1}-\vec{x}_{i}\right) \delta_{{D}}\left(\vec{x}_{2}-\vec{x}_{j}\right) \delta_{{D}}\left(\vec{x}_{3}-\vec{x}_{j}\right)\right\rangle+2 \rm{cyc.}\right] \\
&+\left\langle\sum_{i, j, k} \delta_{{D}}\left(\vec{x}_{1}-\vec{x}_{i}\right) \delta_{{D}}\left(\vec{x}_{2}-\vec{x}_{j}\right) \delta_{{D}}\left(\vec{x}_{3}-\vec{x}_{k}\right)\right\rangle \\
=& \delta_{{D}}\left(\vec{x}_{1}-\vec{x}_{2}\right) \delta_{{D}}\left(\vec{x}_{1}-\vec{x}_{3}\right) \bar{n}+\left[\delta_{{D}}\left(\vec{x}_{2}-\vec{x}_{3}\right) \bar{n}^{2}\left(1+\xi^{(2)}_{12}\right)+2 \mathrm{cyc} .\right]+\bar{n}^{3}\left(1+\xi^{(2)}_{12}+\xi^{(2)}_{23}+\xi^{(2)}_{31}+\xi^{(3)}_{123}\right),
\end{aligned}
\ee
where $\xi^{(3)}_{123}$ is the continuous three-point correlation function. Using Eqs. (\ref{eq:a2}) and (\ref{eq:a6}), we can obtain the galaxy 3-point correlation function for a single tracer:
\be
\xi^{(3)}(\vec x_1, \vec x_2, \vec x_3) = \frac{1}{{\bar n}^2} \delta_{{D}}\left(\vec{x}_{1}-\vec{x}_{2}\right) \delta_{{D}}\left(\vec{x}_{1}-\vec{x}_{3}\right) + \left[ \frac{\delta_D(\vec x_2 -\vec x_3)}{\bar n} \xi^{(2)}_{12} + 2 \rm{cyc.}\right] + \xi^{(3)}.
\ee
The observed galaxy bispectrum is given by
\be
B_{g,\rm measured}(k_1,k_2,k_3) =B_g(k_1,k_2,k_3) +  \frac{1}{{\bar n}^2} + \frac{1}{\bar n}[P_g(k_1)+2{\rm cyc.}] .
\ee
So the shot noise contribution to the skew spectrum is
\be
S_{\rm 1T}(k) = \int \frac{d^{3} \vec{q}}{(2 \pi)^{3}} \left[\frac{1}{\bar n}\left( P_g(k)+ P_g(q)+P_g(\alpha) \right)+\frac{1}{\bar n^2}\right] .
\ee
When we consider two different tracers, for example, the 3-point  correlation function $\left\langle\delta^{[1]}_{g}\left(\vec{x}_{1}\right) \delta^{[1]}_{g}\left(\vec{x}_{2}\right)\delta^{[2]}_{g}\left(\vec{x}_{3}\right)\right\rangle$.
Following the above calculation, the correlation function is given by
\be
\left\langle\delta^{[1]}_{g}\left(\vec{x}_{1}\right) \delta^{[1]}_{g}\left(\vec{x}_{2}\right)\delta^{[2]}_{g}\left(\vec{x}_{3}\right)\right\rangle
 = \frac{1}{\bar n_1}\delta_D(\vec x_1 - \vec x_2) \xi^{(2)}_{23,(1,2)} + \xi^{(3)} ,
\ee
and the corresponding bispectrum is
\be
B_{g,\rm measured}(k_1,k_2,k_3) = \frac{1}{\bar n_1} P_{g,(1,2)}(k_3) + B_g(k_1,k_2,k_3) .
\ee
Finally, we can obtain the shot noise contributions to the skew spectra when we consider two different tracers.
\ba
S^{(s)}_{\rm (11,2)}(k)&=&\int \frac{d^{3} \vec{q}}{(2 \pi)^{3}} \frac{1}{\bar n_1}P_{g,(1,2)}(k) {\rm~~~~~~if~set}~\vec x_2 = \vec x_1,\\
S^{(s)}_{\rm (12,1)}(k)&=&\int \frac{d^{3} \vec{q}}{(2 \pi)^{3}} \frac{1}{\bar n_1}P_{g,(1,2)}(q) {\rm~~~~~~if~set}~\vec x_3= \vec x_1.
\ea

\end{document}